\documentclass[10pt,pra,letterpaper,typeset,twocolumn,showpacs,floatfix]{revtex4-1}

\usepackage{color}
\usepackage{graphicx}
\usepackage{amsmath}
\usepackage{amssymb,amsthm}
\graphicspath{{pict/}{}}

\usepackage{bm}

\newcounter{Fig}

\newcommand{\be}{\begin{equation}}
\newcommand{\ee}{\end{equation}}

\begin{document}

\title{Transport in Sawtooth photonic lattices}

\author{Steffen Weimann}
\affiliation{Institute of Applied Physics, Abbe Center of Photonics, Friedrich-Schiller-Universit\"{a}t Jena, Max-Wien-Platz 1, 07743 Jena, Germany}

\author{Luis Morales-Inostroza}
\affiliation{Departamento de F\'isica and MSI-Nucleus on Advanced
Optics, Center for Optics and Photonics (CEFOP), Facultad de
Ciencias, Universidad de Chile, Santiago, Chile}

\author{Basti\'an Real}
\affiliation{Departamento de F\'isica and MSI-Nucleus on Advanced
Optics, Center for Optics and Photonics (CEFOP), Facultad de
Ciencias, Universidad de Chile, Santiago, Chile}

\author{Camilo Cantillano}
\affiliation{Departamento de F\'isica and MSI-Nucleus on Advanced
Optics, Center for Optics and Photonics (CEFOP), Facultad de
Ciencias, Universidad de Chile, Santiago, Chile}

\author{Alexander Szameit}
\affiliation{Institute of Applied Physics, Abbe Center of Photonics, Friedrich-Schiller-Universit\"{a}t Jena, Max-Wien-Platz 1, 07743 Jena, Germany}

\author{Rodrigo A. Vicencio}
\affiliation{Departamento de F\'isica and MSI-Nucleus on Advanced
Optics, Center for Optics and Photonics (CEFOP), Facultad de
Ciencias, Universidad de Chile, Santiago, Chile}





\begin{abstract}
We investigate, theoretically and experimentally, a photonic realization of a Sawtooth lattice. This special lattice exhibits two spectral bands, with one of them experiencing a complete collapse to a highly degenerate flat band for a special set of inter-site coupling constants. We report the observation of different transport regimes, including strong transport inhibition due to the appearance of the non-diffractive flat band. Moreover, we excite localized Shockley surfaces states, residing in the gap between the two linear bands.
\end{abstract}


\maketitle

Commonly, lattices of coupled waveguides exhibit a dispersive band structure, such that different eigenmodes acquire different phases during evolution. As a consequence, excited waves in general diffract. This effect lies at the heart of the conduction properties of any periodic material. Although the propagation of waves in such photonic lattices is well understood~\cite{rep1,rep2}, some lattices have become prominent for exhibiting peculiar non-diffractive properties due to particular characteristics of their spectrum. These structures are perfectly periodic and yet exhibit insulating properties, which is in contrast to Anderson localization~\cite{ande1,ande2,ande3}, where transport is inhibited by perturbing the underlying periodicity. The non-intuitive transport behavior in these lattices arises from the fact that at least one of their bands is completely flat, that is, all eigenmodes forming this band are degenerate. These extended Bloch modes can be coherently superposed, forming highly localized Flat band states with a strictly zero background~\cite{berg}. As a consequence, the observable transport regimes are very sensitive to how the unit cell is excited~\cite{lieb1}. Aside from the fundamental interest on the transport properties of periodic systems, flat band lattices are a very promising candidate for non-diffractive image propagation~\cite{kag2,lieb3}.
%
\begin{figure}[t]
\centering
\includegraphics[width=0.46\textwidth]{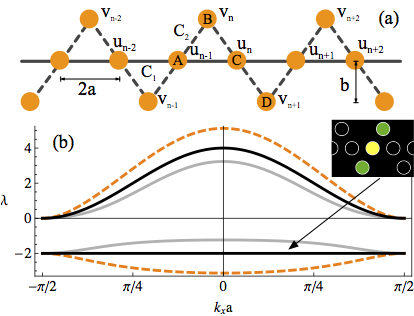}
\caption{(Color online) (a) Implementation of a Sawtooth waveguide lattice. The inter-site coupling $C_1$ ($C_2$) is represented by full (dashed) straight lines. (b) Linear spectrum of an infinite Sawtooth lattice for $C_2 = C_1$ (gray), $C_2 = \sqrt{2}C_1$ (black), and $C_2 = 2C_1$ (dashed). ($C_1=1$). The inset in (b) shows the profile of a Sawtooth Flat band mode (black, green and yellow correspond to intensities $0,\ 1,\ 2$, respectively).}\label{fig1}
\end{figure}
%
Well-known two-dimensional (2D) examples of non-diffractive lattices are Kagome and Lieb lattices. Recently, the experimental implementation of Kagome lattice~\cite{BRD11, kag1} and Lieb lattices~\cite{lieb1} were studied, showing that transport strongly depends on which site in the unit cell is excited. Fundamental flat band modes in Lieb lattices were reported in Refs.~\cite{lieb3,lieb4,chenlieb}. Since flat band modes do not diffract across the lattice, any linear combination of them propagates without diffraction as well~\cite{berg,kag2}. Additionally, a very recent work explores the construction of a full flat band using a band engineering method in a photonic crystal, going beyond the tight-binding limit~\cite{SRfullKFB}. Although the existence of flat bands does not require a 2D lattice~\cite{stubMario, PREsaw, PREkagrib}, there are only few experimental realizations of flat bands in quasi one-dimensional (1D) lattices in general. Very recently, the observation of a linear localized state occupying two sites only in a rhombic (diamond) configuration was reported~\cite{diamondThomson}. In the context of micro-pillar optical cavities, polariton condensation was demonstrated experimentally in 1D Lieb (Stub) lattices~\cite{stubAmo}. A peculiar incarnation of a 1D system possessing a flat band is the Sawtooth lattice~\cite{saw1} (also known as $\Delta$ chain~\cite{saw2}), used to describe the properties of $YCuO_{2.5}$ conducting delafossites~\cite{Cu}. The Sawtooth lattice consists of a 1D sequence of triangles forming a lattice; a sketch of it is shown in Fig.~\ref{fig1}(a). Note that we show a particular implementation where the triangles have alternating orientation in order to suppress the coupling between successive vertices $B$ and $D$~\cite{PREsaw}. Recent studies on the Sawtooth lattice include quantum topological excitations~\cite{QTE} as well as Bose-Einstein condensation in flat band systems~\cite{huber}, with a recent proposal for an experimental implementation~\cite{SRbecsaw}. However, a realization of this type of lattice in the laboratory is elusive so far.

In our work, we implement an artificial Sawtooth lattice using a coupled waveguide structure and investigate its transport characteristics. We explore the transition into localization for different parameters of the Sawtooth geometry and demonstrate the predicted flat band properties. In particular, we show the existence of Shockley-type edge states~\cite{shockley}, as exact solutions at $A$-site edges.

The unit cell of the Sawtooth lattice is not affected by the orientation of the triangles as the underlying symmetry is manifested in the tight-binding Hamiltonian: The unit cell consists of only two elements, $A$ and $B$. In this binary super-lattice, the evolution of the electric field amplitude along the propagation direction $z$ is well described by a sequence of coupled Schr\"odinger equations:
\begin{eqnarray}
-i\frac{d u_n}{d z}&=&C_1\left(u_{n+1}+u_{n-1}\right)+C_2\left(v_{n}+v_{n+1}\right)\ ,\nonumber\\
-i\frac{d v_n}{d z}&=&C_2\left(u_{n}+u_{n-1}\right)\ ,\
\label{dls}
\end{eqnarray}
where $u_{n}$ and $v_{n}$ represent the electric field amplitudes at the $A$ and $B$ sites, respectively, according to Fig.~\ref{fig1}(a). The coupling between two $A$ sites is defined by $C_1$ whereas coupling between $A$ and $B$ sites is denoted $C_2$ [visualized by full and dashed lines, respectively, in Fig.~\ref{fig1}(a)]. When implementing this lattice as an array of coupled waveguides, the strength of the coupling constants $C_1$ and $C_2$ follows an exponential decaying law on the distance between lattice sites~\cite{expo}. For simplicity, we define the ratio $\delta\equiv C_2/C_1$ to describe different regimes of the transport on the Sawtooth lattice. In the geometry of the lattice, we set the separation between successive $A$ sites as $2a$, and the vertical distance between $A$ and $B$ sites as $b$.
%
%
\begin{figure}[t]
\centering
\includegraphics[width=0.47\textwidth]{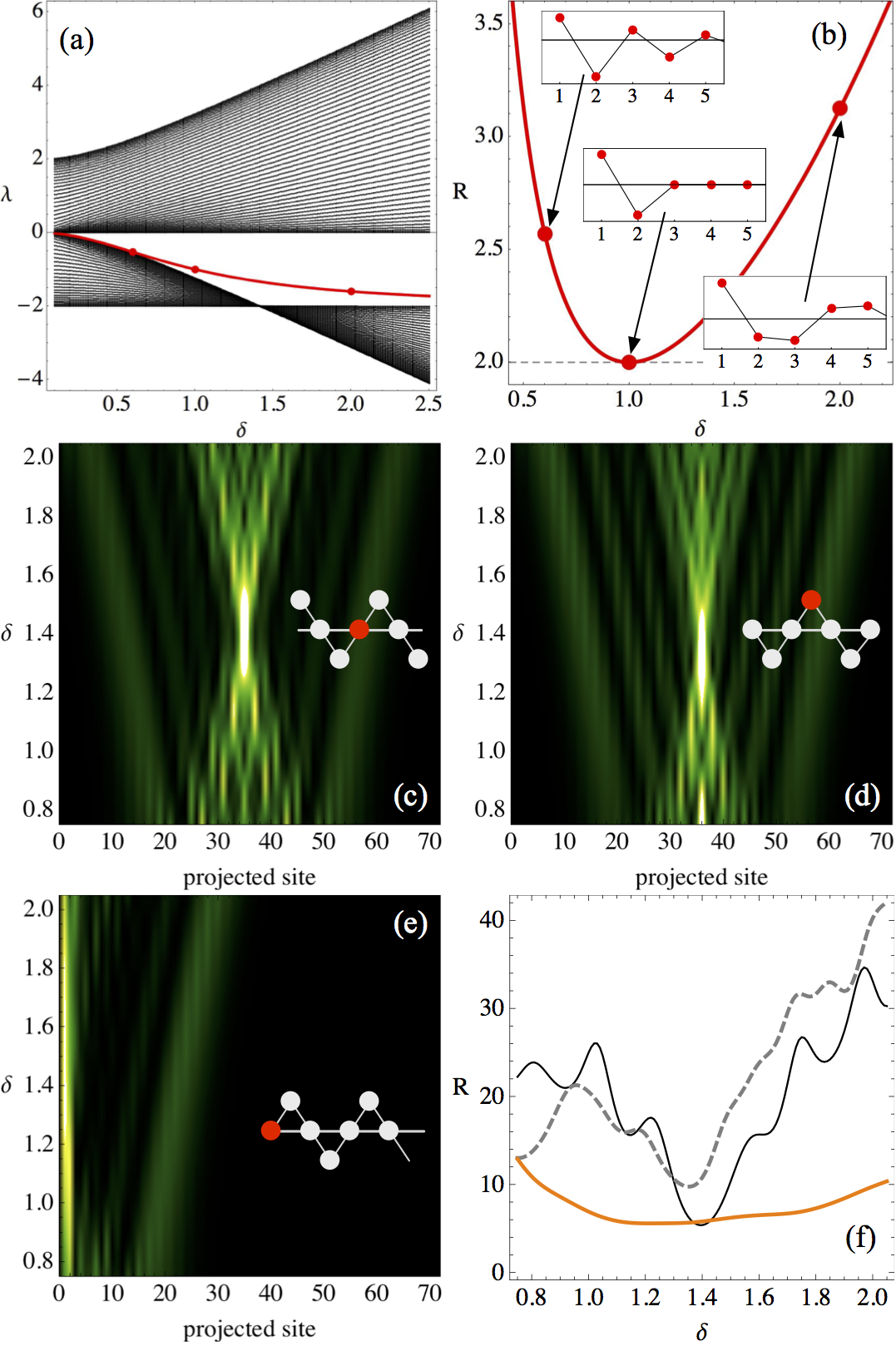}
\caption{(Color online) (a) Linear spectrum $\lambda$ vs. $\delta$. The edge state appears in red. (b) The participation ratio $R$ versus $\delta$ for surface modes. [Red dots in (a) and (b) correspond to profiles shown as insets in (b)]. Numerically obtained output intensity profiles versus $\delta$ for single-site excitation at the (c) bulk $A$-site, (d) bulk $B$-site, (e) edge $A$-site. (f) Output $R$ versus $\delta$ for bulk-$A$ (thin), bulk-$B$ (dashed), and edge-$A$ (orange) sites. ($C_1=1$).}\label{fig2}
\end{figure}
%
The plane wave solutions of \eqref{dls} are of the form $\{u_{n}(z),v_{n}(z)\}=\{U,V\} \exp{(i k_x x_n)}\exp{(i \lambda z)}$, where $U$ and $V$ describe constant amplitudes. $k_x$ defines the transverse wavenumber, and $x_n=a n$ determines the horizontal position of the $A$ and $B$ sites. The dispersion relation between the longitudinal spatial frequency $\lambda$ and $k_x$ follows as
\begin{equation*}
\lambda(k_x)=C_1\left[\cos(2k_x a)\pm \sqrt{1+4(\delta^2-1)\cos^2(k_x a)+4\cos^4(k_x a)}\right]\ ,
\end{equation*}
which is the spectrum of the Sawtooth lattice having two bands of different curvature and width. Figure~\ref{fig1}(b) shows the linear spectrum of a Sawtooth lattice, in the first Brillouin zone, for three representative values of $\delta$. We see that for a critical value $\delta_c\equiv\sqrt{2}$, the bands reduce to $\lambda(k_x)=4C_1 \cos^2 (k_x a)$ and $\lambda(k_x)=-2C_1$; i.e., the lower band collapses and becomes completely flat [see black line in Fig.~\ref{fig1}(b)]. The states residing in this non-dispersive and highly degenerated band spread only across three sites, namely $\{...,0,-1,\sqrt{2},-1,0,...\}$, as Fig.~\ref{fig1}(b)-inset shows. These localized states can be located anywhere in the lattice. As they possess the same propagation constant, they are promising candidates for non-diffractive image propagation~\cite{kag2, lieb3}. For $\delta<\delta_c$, the sign of the curvature is the same for both bands, such that at a given $k$ the states from both bands will propagate in the same direction (but with different velocities). For $\delta>\delta_c$, however, the curvature is opposite, and the states at a fixed $k$ propagate in opposite directions. Hence, at the critical value $\delta_c$ the sign of the curvature of the lower band changes, which is inherent to Sawtooth lattices. One possible application of this phenomenon is to use this type of lattice as a beam splitter, with two beams traveling in the same or in opposite directions.

Additionally, we numerically compute the spectrum of a finite Sawtooth lattice by directly diagonalizing~\eqref{dls}, for different values of $\delta$, and obtain the spectrum shown in Fig.~\ref{fig2}(a). The analytically computed bands $\lambda(k_x)$ agree perfectly with these numerical results but, additionally, an edge state appears [see red curve in Fig.~\ref{fig2}(a)]. This mode exists only at the $A$-site edge, and for two $A$-ends two degenerate states appear. The mode is well localized and decays into the bulk in an exponential way [see examples in Fig.~\ref{fig2}(b)]. Using the ansatz
$$\{u_{n}^e(z),v_{n}^e(z)\}=\{A,B\}\ \epsilon^{n}\exp{(i \lambda_e z)}\ ,$$
with $n\geqslant 0$ at the $A$-site edge and $|\epsilon|<1$. We analytically find that $B=-A/\delta$, $\epsilon = (1-\delta^2)/(1+\delta^2)$ and $\lambda_e=-2C_1\delta^2/(1+\delta^2)$. This edge solution coincides perfectly with the numerically found edge modes. To study the effective size of the edge state and its dependence on the parameter $\delta$, we use the participation ratio, defined as $R\equiv (\sum_{n}|w_{n}|^2)^2/\sum_{n}|w_{n}|^4$, where $w_n$ is the field amplitude of the edge state at site $n$. In Fig.~\ref{fig2}(b), we plot the corresponding function $R(\delta)=(1+\delta^2)^2/2\delta^2$ for the edge state. We find a minimum of $R=2$ for $\delta=1$. At this point, $\epsilon=0$ and the mode occupies only the first two lattice sites at the edge with equal amplitude but opposite phases. This highly localized state shows a fundamental condition found in diverse flat band systems, namely that the coupling between these two sites with respect to a third one is completely canceled, and the transmission of energy through the rest of the lattice is forbidden. For $\delta\rightarrow 0$, the decay factor $\epsilon\rightarrow 1$, hence, the participation ratio increases rapidly showing the transition of the mode into an extended mode inside the band. Consequently, for $\delta=0$, the $A$ and $B$ layers are completely decoupled from each other and no surface state exists~\cite{rep1,rep2,surf1,surf2}. For $\delta>1$, $R$ increases slowly as $\epsilon\rightarrow -1$. In this regime, the surface state is still single-peaked but acquires a more complicated phase structure, namely $0,\pi,\pi,0,0,...$ [see inset in Fig.~\ref{fig2}(b)]. For $\delta\gg 1$, the surface mode converges to the upper mode of the lower band at $\lambda_e = -2C_1$.

To study general transport properties, we explore the excitation of individual waveguides at the bulk and surface of a Sawtooth lattice. Figs.~\ref{fig2}(c)--(e) show the output intensity profile for one-site input excitations, obtained by numerically integrating~\eqref{dls}, up to the same propagation distance. Figure~\ref{fig2}(c) shows the tendency to localization for $\delta\rightarrow \delta_c$ for an $A$-site bulk excitation. It is important to note that a single-site input excites both bands homogeneously. However, faster waves, mostly coming from the upper band, are hardly visible due to the larger spreading area. Slower waves, mostly belonging to the lower band, spread over a narrower region and possess larger site amplitudes. Close to $\delta_c$, mostly a flat band state is excited, such that the output profile is very narrow with only a weak background arising from the weak excitation of modes from the upper band. For $\delta>\delta_c$, the diffraction is stronger, because the lower band is not flat anymore. For $\delta<\delta_c$, both bands also consist of extended modes, but as the band curvature is reduced, the diffraction cone is slightly weaker. In Fig.~\ref{fig2}(f) it is shown that participation ratio of the output light distribution reaches a minimum when $\delta\approx\delta_c$. For a $B$-site bulk excitation [see figure~\ref{fig2}(d)], the observation is quite similar but, additionally, a tendency to generate a localized pattern for lower $\delta$ is observed. This can be well understood by considering the fact that the underlying reduction of the diagonal coupling $C_2$ gradually isolates the top or bottom sites from the rest of the lattice. In Fig.~\ref{fig2}(f) we observe how the output participation ratio decreases for $\delta \approx\delta_c$ and for $\delta\approx 0.75$, and grows for $\delta>\delta_c$. This $\delta$-controlled bulk transport transition can be viewed as an insulator-conductor system governed by the geometry of the lattice. Additionally, these results nicely show how Sawtooth lattices are able to localize energy only by virtue of their geometry, without requiring, e.g., disorder or nonlinearity~\cite{rep1,rep2}.

When exciting the $A$-site edge, we observe a stable and localized intensity distribution at the surface [see Fig.~\ref{fig2}(e)]. This is confirmed by the participation ratio shown in Fig.~\ref{fig2}(f), which exhibits a minimum for $\delta\approx 1.1$. We observe that for $\delta\approx 1$ the profile is rather flat at the surface with essentially two equal peaks ($B=-A$), which is in agreement with the modes shown in Fig.~\ref{fig2}(b). For $\delta\gtrsim 1$ the excited profile becomes single-peaked at the edge ($|B|<|A|$), and the effective size tends to be constant, but slowly increasing as $\delta$ increases. For $\delta\lesssim 1$, the localized profile shifts its center to the second waveguide ($|B|>|A|$) and the participation ratio increases.
\begin{figure}[t]
\centering
\includegraphics[width=0.48\textwidth]{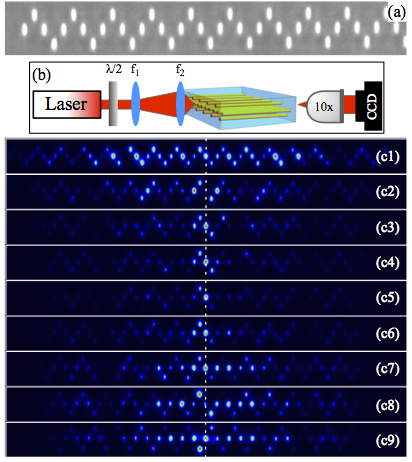}
\caption{(Color online) (a) Microscope image at the output facet of a Sawtooth photonic lattice. (b) Experimental setup for studying transport in this lattice. (c1)--(c8) Experimental output intensity for an $A$-site bulk excitation with $\delta=1.89,\ 1.68,\ 1.49,\ 1.44,\ 1.38,\ 1.35,\ 1.13,\ 1.05,\ 0.93$, respectively. The dashed line indicates the input position.}\label{fig3}
\end{figure}

We fabricated several waveguide lattices in the Sawtooth geometry using the direct femtosecond laser-writing technique~\cite{fslt}, on a $L=10\, cm$ long fused silica glass chip. A microscopic image of a fabricated lattice is shown in Fig.~\ref{fig3}(a). Each fabricated array has a constant horizontal period of $a=11\, \mu m$, but different vertical distances in the interval $b\in \{13, 20\}\, \mu m$, such that the ratio $\delta$ decreases from $1.89$ to $0.82$. This parameter was calibrated varying the vertical distance $b$ by directly measuring the propagation pattern in a $3$-sites triangular array, while keeping constant the horizontal distance $a$. Moreover, the arrays consist of $71$ waveguides each, with $A$-site edges, as shown in Fig.~\ref{fig3}(a). To study transport in this lattice, we inject a horizontally polarized $633\, nm$ laser beam into single waveguides, by tightly focusing the beam with a lens of short focal distance. We use a $10\times$ microscope objective to image the intensity patterns from the output facet onto a CCD camera [see Fig.~\ref{fig3}(b)]. The exposition of our camera is set such that the background is observable as well.

We first study bulk transport by exciting a central $A$ site in each lattice, as presented in Fig.~\ref{fig3}(c). We observe how transport is governed by the properties of the two bands discussed above. Similar to our simulations, faster modes residing in the upper band are weakly visible compared to the slower modes from the lower band. For a smaller value of $b$ (i.e., larger $\delta$) light strongly spreads all over the lattice, as show in Figs.~\ref{fig3}(c1) and (c2). When the vertical distance is increased and $\delta$ approaches $\delta_c$ [see Figs.~\ref{fig3}(c3)--(c6)], we observe light trapping around the excited waveguide. This is a strong indication of the reduction of the curvature and width of the lower band, which is the most excited band for this localized input condition (a single-site excitation is very close to a three-sites flat band mode profile). Taking a look at the background of the intensity distributions, we observe that the upper band, where the modes possess a flat phase distribution, is excited only weakly and some light propagates away from the input excitation. For $\delta<\delta_c$, the diffraction of the background increases again, but is weaker compared to the case where $\delta>\delta_c$, such that the light is concentrated around the excited site with a more homogenous amplitude profile [see Figs.~\ref{fig3}(c7)--(c9)]. All these observations agree very well with our numerical results shown in Fig.~\ref{fig2}(c).

\begin{figure}[h]
\centering
\includegraphics[width=0.485\textwidth]{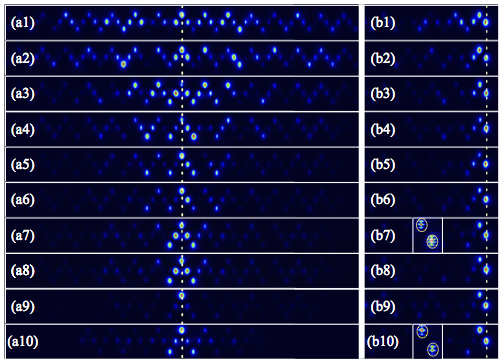}
\caption{(Color online) Experimental output intensity for a (a) $B$-site bulk and for an (b) $A$-site edge excitation. Insets in (b7) and (b10) show an interferogram between the respective output profile and a tilted plane wave. $\delta=1.89,\ 1.68,\ 1.49,\ 1.44,\ 1.38,\ 1.35,\ 1.13,\ 1.05,\ 0.93,\ 0.82$, respectively. Dashed line indicates input position.}\label{fig4}
\end{figure}

In the next experiment we excite a bulk $B$ site in each lattice, as shown in Fig.~\ref{fig4}(a). We observe that for $\delta>\delta_c$ the light spreads slower than for a bulk $A$ excitation, but nevertheless a good transport over the lattice is observed. Again, there is a tendency to localization close to $\delta_c$, but not as strong as in the previous case where an $A$ site was excited. Notably, the strongest localization is obtained for a value of  $\delta=0.93$, that is, slightly off the critical value, which is in agreement with the numerical simulations presented in Fig.~\ref{fig2}(d).
Finally, we explore the excitation of edge states by injecting light at the $A$-site edge. Our results are shown in Fig.~\ref{fig4}(b). In general, for any $\delta$-value, we observe the excitation of a very well localized pattern, as expected from our simulations  shown  in Fig.~\ref{fig2}(e)]. As described above, we observe a transition from a two-site localized profile (for $\delta\approx 1$) to a one-peaked profile for increasing $\delta$. The light diffraction pattern essentially corresponds to the excitation of the predicted surface state discussed in Figs.~\ref{fig2}(a) and (b). We include two insets in Fig.~\ref{fig4}(b) to show the phase structure of this profile for the first $A$ and $B$ sites, observing a clear $\pi$-phase difference between them. Following the formal definitions, this surface profile corresponds to an edge state originated by an effective defect at the surface that arises from the reduction of nearest-neighbor interactions~\cite{plotnik,lieb1} as well as from a band crossing occurring at $\delta=\delta_c$, as shown in Fig.~\ref{fig2}(a). By inspecting the states in the lower band, we observe a kind of band twist at this critical value, such that our observed edge state is of the Shockley-type~\cite{chen1,chen2}.

In conclusion, we have studied the fundamental transport properties of a Sawtooth photonic lattice. We have shown that bulk transport strongly depends on the particular geometry of this lattice, with a strong tendency to localization for $\delta\approx\delta_c$. For $A$- or $B$-site bulk excitation, the energy tends to concentrate strongly at the input region and transport is reduced drastically. We found that for top or bottom bulk excitation, light tends to localize even for smaller values of $\delta$, depending on the particular propagation distance. When injecting light at the $A$-edge, a localized Shockley surface state is excited. Therefore, in general, this special lattice allows us to localize energy in different positions, depending on their particular coupling parameters, beyond the excitation of the flat band phenomenology. This is very important when considering the use of waveguide lattices for transmitting information in the low-power regime, that is, without employing nonlinearities.

This work was supported in part by Programa ICM grant RC130001, FONDECYT Grant No. 1151444, the Deutsche Forschungsgemeinschaft (grant NO 462/6-1, SZ 276/7-1, SZ 276/9-1, BL 574/13-1), and the German Ministry of Education and Research (Center for Innovation Competence program, grant 03Z1HN31).


\begin{thebibliography}{99}

\bibitem{rep1} F. Lederer, G.I. Stegeman, D.N. Christodoulides, G. Assanto, M. Segev, and Y. Silberberg, Phys. Rep. \textbf{463}, 1 (2008).

\bibitem{rep2}S. Flach and A. Gorbach, Phys. Rep. {\bf 467}, 1 (2008).

\bibitem{ande1}P.W. Anderson, Phys. Rev. {\bf 109}, 1492 (1958).

\bibitem{ande2}T. Schwartz, G. Bartal, S. Fishman, and M. Segev, Nature {\bf 446}, 52 (2007).

\bibitem{ande3}Y. Lahini, A. Avidan, F. Pozzi, M. Sorel, R. Morandotti, D. N. Christodoulides, and Y. Silberberg, Phys. Rev. Lett. {\bf 100}, 013906 (2008).

\bibitem{berg}D.L. Bergman, C. Wu, and L. Balents, Phys. Rev. B \textbf{78}, 125104 (2008).

\bibitem{lieb1}D Guzm\'an-Silva, C. Mej\'ia-Cort\'es, M.A. Bandres, M.C. Rechtsman, S. Weimann, S. Nolte, M. Segev, A. Szameit and R.A. Vicencio, New J. Phys. \text{16}, 063061 (2014).


\bibitem{kag2}R.A. Vicencio and C. Mej\'ia-Cort\'es, J. Opt. \textbf{16}, 015706 (2014).

\bibitem{lieb3}R.A. Vicencio, C. Cantillano, L. Morales-Inostroza, B. Real, C. Mej\'ia-Cort\'es, S. Weimann, A. Szameit, and M.I. Molina, Phys. Rev. Lett. \textbf{114}, 245503 (2015).

\bibitem{BRD11} M. Boguslawski, P. Rose, and C. Denz, Appl. Phys. Lett. {\bf 98}, 061111 (2011).

\bibitem{kag1}R.A. Vicencio and M. Johansson, Phys. Rev. A \textbf{87} 061803(R) (2013).


\bibitem{lieb4}S. Mukherjee, A. Spracklen, D. Choudhury, N. Goldman, P. \"{O}hberg, E. Andersson, and R.R. Thomson, Phys. Rev. Lett. \textbf{114}, 245504 (2015).

\bibitem{chenlieb}S. Xia, Y. Hu, D. Song, Y. Zong, L. Tang, and Z. Chen, Opt. Lett. \textbf{41}, 1435 (2016).

\bibitem{SRfullKFB}C. Xu, G. Wang, Z.H. Hang, J. Luo, C.T. Chan, and Y. Lai, Sci. Rep. \textbf{5}, 18181 (2015).

\bibitem{stubMario} M.I. Molina, Phys. Rev. A \textbf{92}, 063813 (2015).

\bibitem{PREsaw}M. Johansson, U. Naether, and R.A. Vicencio, Phys. Rev. E \textbf{92}, 032912 (2015).

\bibitem{PREkagrib}P.P. Beli\v{c}ev, G. Glirori\'c, A. Radosavljevi\'c, A. Maluckov, M. Stepi\'c, R.A. Vicencio, and M. Johansson, Phys. Rev. E \textbf{92}, 052916 (2015).

\bibitem{diamondThomson}S. Mukherjee and R.R. Thomson, Opt. Lett. {\bf 40}, 5443 (2015).

\bibitem{stubAmo}F. Baboux, L. Ge, T. Jacqmin, M. Biondi, A. Lema\^{i}tre, L. Le Gratiet, I. Sagnes, S. Schmidt, H.E. T\"ureci, A. Amo, J. Bloch, Phys. Rev. Lett. \textbf{116}, 066402 (2016).

\bibitem{saw1}T. Nakamura and K. Kubo, Phys. Rev. B \textbf{53}, 6393 (1996).

\bibitem{saw2}D. Sen, B.S. Shastry, R.E. Walstedt, and R. Cava, Phys. Rev. B \textbf{53}, 6401 (1996).

\bibitem{Cu}R.E. Walstedt, R.J. Cava, R.F. Bell, J.J. Krajewski, and W.F. Peck, Jr., Phys. Rev. B \textbf{49}, 12369 (R) (1994).

\bibitem{QTE}S.A. Blundell and M.D. N\'u\~nez-Regueiro, Eur. Phys. J. B \textbf{31}, 453 (2003).

\bibitem{huber}S.D. Huber and E. Altman, Phys. Rev. B \textbf{82}, 184502 (2010).

\bibitem{SRbecsaw}T. Zhang and G.-B. Jo, Sci. Rep. \textbf{5}, 16044 (2015).

\bibitem{shockley}W. Shockley, Phys. Rev. \textbf{56}, 317 (1939).

\bibitem{expo}A. Szameit, F. Dreisow, T. Pertsch, S. Nolte, and A. T\"unnermann, Opt. Exp. \textbf{15}, 1579 (2007).


\bibitem{surf1}C.R. Rosberg, D.N. Neshev, W. Krolikowski, A. Mitchell, R.A. Vicencio, M.I. Molina, and Yu.S. Kivshar, Phys. Rev. Lett. {\bf 97}, 083901 (2006).

\bibitem{surf2} E. Smirnov, M. Stepi\'c, C.E. R\"uter, D. Kip, and V. Shandarov, Opt. Lett. \textbf{31}, 2338 (2006).

\bibitem{fslt}A. Szameit and S. Nolte S, J. Phys. B: At. Mol. Opt. Phys. \textbf{43}, 163001 (2010).

\bibitem{plotnik}Y. Plotnik, M. C. Rechtsman, D. Song, M. Heinrich, J. M. Zeuner, S. Nolte, Y. Lumer, N. Malkova, J. Xu, A. Szameit, Z. Chen and M. Segev, Nat. Mater. \textbf{13}, 57-62 (2014).

\bibitem{chen1}N. Malkova, I. Hromada, X. Wang, G. Bryant, and Z. Chen, Opt. Lett. \textbf{34}, 1633 (2009); 

\bibitem{chen2}N. Malkova, I. Hromada, X. Wang, G. Bryant, and Z. Chen,Phys. Rev. A \textbf{80}, 043806 (2009).


\end{thebibliography}
\end{document}